\documentclass[conference]{IEEEtran}
\usepackage{amssymb}
\usepackage{amsfonts}
\usepackage{times,amsmath,epsfig}
\usepackage{color}
\usepackage{latexsym}
\begin{document}
\title{\huge Iterative Detection and Decoding for MIMO Systems with Knowledge-Aided
Message Passing Algorithms \vspace{-0.25em}}

\author{Jingjing~Liu, Peng Li and Rodrigo~C.~de~Lamare \\
\IEEEauthorblockA{Department of Electronics, The University of York\\
Heslington, York, YO10 5DD, UK \\
Email: \{jl622; pl534; rcdl500\}@york.ac.uk}\vspace{-0.25em}}
%\author{Jingjing Liu, Peng Li and Rodrigo C. de Lamare \vspace{-2em}
%\thanks{Jingjing Liu and Rodrigo C. de Lamare \{jl622;rcdl500\}@ohm.york.ac.uk are with Department of Electronics, The University of York, Heslington, York, YO10 5DD, UK. The authors are grateful to Henk Wymeersch who provided MATLAB codes of URW-BP algorithm for comparison purpose.}}% <-this % stops a s

\maketitle

%\IEEEpeerreviewmaketitle

\begin{abstract}

In this paper, we consider the problem of iterative detection and
decoding (IDD) for multi-antenna systems using low-density
parity-check (LDPC) codes. The proposed IDD system consists of a
soft-input soft-output parallel interference (PIC) cancellation
scheme with linear minimum mean-square error (MMSE) receive filters
and two novel belief propagation (BP) decoding algorithms. The
proposed BP algorithms exploit the knowledge of short cycles in the
graph structure and the reweighting factors derived from the
hypergraph's expansion. Simulation results show that when used to
perform IDD for multi-antenna systems both proposed BP decoding
algorithms can consistently outperform existing BP techniques with a
small number of decoding iterations.
\end{abstract}

\vspace{-0.05em}

\section{Introduction}

Multi-input and multi-output (MIMO) systems can support several
independent data streams, resulting in a significant increase of the
system capacity \cite{Telatar}. In order to separate the data
streams and mitigate the interference between them, a detection
algorithm must be employed at the receiver. In the last decade or
so, a great deal of effort has been devoted to the development of
detection algorithms and their integration with channel decoding
techniques \cite{Wang}-\cite{Peng2}. In this context, MIMO systems
with joint detection/decoding have been shown to produce excellent
results, approaching the performance of an interference free
scenario. In a system with joint detection/decoding an ideal
receiver is comprised of two components: an efficient soft-input
soft-output (SISO) MIMO signal detector and a SISO decoder with low
delay. Specifically, the estimated log likelihood ratios associated
with the encoded bits are computed by the detector and these
estimates will serve as input to the decoder. Then in the second
phase of the detection/decoding iteration, the decoder generates
\textit{a posteriori} probabilities for encoded bits of each data
stream. As a result, the soft estimate of the transmitted symbol is
obtained which can facilitate the detection in the first phase of
the next outer iteration. The joint process of detection/decoding is
then repeated in an iterative manner until the maximum number of
iterations is reached. However, in practice there are many open
issues for such an IDD scheme, e.g. severe detection/decoding delay
especially for codes with short block lengths \cite{Brink2},
\cite{Hou}, or prohibitively high computational complexity
associated with IDD systems in general.

Low-density parity-check (LDPC) codes, invented by Gallager
\cite{Gallager} are a class of linear block codes which can achieve
near-Shannon capacity with linear-time encoding and parallelizable
decoding algorithms. The standard BP algorithm is well-known as the
most effective algorithm to decode LDPC codes \cite{RyanLin}, and
has been widely employed as part of IDD schemes for MIMO systems
\cite{Brink2}, \cite{Wu} and \cite{Ding}. It can produce exact
inference solutions only if the graphical model does not contain
short cycles. With the existence of cycles, the standard BP
algorithm has a number of shortcomings, such as convergence to a
codeword is not guaranteed and convergence to a codeword can take
many iterations, especially at low signal to noise ratios (SNR),
which significantly deteriorate the decoding performance and cause
unexpected transmission delay. Due to this fact, many applications
of LDPC-coded MIMO systems have a performance degradation at some
extent. In \cite{Wainright2}, the authors converted the problem of
finding the fixed points of BP algorithms into that of solving a
variational problem, and defined a set of reweighting factors.
Recently, Wymeersch et al. \cite{Wymeersch2} extended the use of
reweighted BP algorithm from pairwise graphs to hypergraphs and
reduced the set of reweighted parameters to a constant value,
whereas Liu and de Lamare \cite{vfap} considered two possible
values.

In this paper, we develop an efficient IDD scheme for MIMO systems
operating in a spatial multiplexing configuration with a reduced
complexity and a low delay. The proposed scheme consists of a SISO
parallel interference cancellation (PIC) scheme with linear minimum
mean-square error (MMSE) receive filters and two novel
knowledge-aided (KA) belief propagation (BP) decoding algorithms.
The first KA decoding algorithm is termed cycles knowledge-aided
reweighted BP (CKAR-BP) algorithm, whereas the second KA decoding
techniques is called expansion knowledge-aided reweighted BP
(EKAR-BP) algorithm. In the following, we present an IDD scheme for
MIMO systems equipped with the proposed KA BP algorithms which can
considerably improve the performance of existing schemes. The
proposed CKAR-BP decoder takes advantage of the cycle distribution
of the Tanner graph, while the proposed EKAR-BP decoder first
expands the original graph into a number of subgraphs then locally
optimizes the reweighting parameters. Incorporated with a SISO
PIC-MMSE detector, both CKAR-BP and EKAR-BP algorithms are shown to
outperform the standard BP and the uniformly reweighted BP (URW-BP)
\cite{Wymeersch2} algorithms when performing IDD for MIMO systems.

The organization of this paper is as follows: Section
\uppercase\expandafter{\romannumeral 2} introduces the system model. In
Section \uppercase\expandafter{\romannumeral 3}, the proposed
EKAR-BP and CKAR-BP algorithms are explained in detail. Section
\uppercase\expandafter{\romannumeral 4} shows the simulation results
along with discussions. Finally, Section
\uppercase\expandafter{\romannumeral 5} concludes the paper.

\section{System Model}

Let us consider a narrowband  MIMO system with $N_T$ transmit
antennas and $N_R$ receive antennas ($N_R \geq N_T$). The MIMO
system operates in a spatial multiplexing configuration and
transmits data over flat fading channels. The received data  after
demodulation, matched filtering and sampling is collected in a
vector $\boldsymbol {r} \in \mathbb{C}^{N_R \times 1}$ with
sufficient statistics for detection and given by
\begin{equation}\label{1}
\boldsymbol {r} = \boldsymbol{C}\boldsymbol{s} + \boldsymbol{n},
\end{equation}
where $\boldsymbol{C} \in \mathbb{C}^{N_R \times N_T}$ is the
channel matrix, $\boldsymbol{s} \in \mathbb{C}^{N_T \times 1}$ is
the encoded data vector and $\boldsymbol{{n}} \in \mathbb{C}^{N_R
\times 1}$ is the noise vector with zero mean and power $\sigma_n^2$
elements. In what follows, we assume that the receiver has perfect
knowledge of the channel matrix $\boldsymbol{C}$. In practice, an
estimation algorithm must be employed to compute the parameters of
$\boldsymbol{C}$ \cite{jio_mimo,Peng2}.

\subsection{PIC-MMSE Detection Algorithm}

In a SISO PIC-MMSE detection algorithm, the estimates of the
transmitted symbols are obtained based on the \textit{a priori}
log-likelihood ratios (LLRs) obtained from the LDPC channel decoder.
These ``soft'' estimates are extracted from the received vector to
perform interference cancellation for a MIMO system. The remaining
noise-plus remaining interference terms are then equalized by a
linear MMSE receive filter which is followed by the computation of
the a posteriori LLRs of the individual constituent bits. The SISO
PIC-MMSE algorithm used as an outer component is detailed in the
following.

According to the SISO model in \cite{Wang}, when processing the
$k$th stream, a PIC detector cancels the interference of all other
streams ($q \neq k$) such that
\begin{equation} \label{2}
\centering {\boldsymbol {\hat{r}}_k=\boldsymbol {r}-\sum_{q \neq k}
\boldsymbol {c}_q\hat{y}_q=\boldsymbol {c}_ks_k+\boldsymbol
{\tilde{n}},~~~\forall k}
\end{equation}
where $y_q, q \neq k$ are the estimates of the transmitted
co-channel symbols obtained from the channel decoder which are
computed according to $\hat{y}_q = E[y_q] = \sum_{a \in \mathcal{O}}
P[y_q = a]a$ where $P[y_q = a]$ corresponds to the a priori
probability of the symbol $a$ on the constellation map
$\mathcal{O}$. The term $\boldsymbol {c}_k$ is the $k$th column of
the channel matrix $\boldsymbol{C}$ and $\boldsymbol {\tilde{n}}$ is
the noise-plus-remaining-interference vector to be equalized by
linear MMSE receive filters as
\begin{equation} \label{3}
\centering \hat{y}_k =
{\boldsymbol{\tilde{w}}_k^H\boldsymbol{\hat{r}}_k=\boldsymbol{\tilde{w}}_k^H\boldsymbol
{c}_ks_k+\boldsymbol{\tilde{w}}_k^H\boldsymbol {\tilde{n}}},
\end{equation}
in which `$(\cdot)^H$' denotes the Hermitian transpose and the MMSE
receive filter is given by
${\boldsymbol{\tilde{w}}_k^H=E_s\boldsymbol{{c}}_k^H\big{(}\boldsymbol{C}\tilde{\boldsymbol{\Lambda}_k}\boldsymbol{C}^H+N_0\boldsymbol{I}_{N_R}\big{)}},$
where $E_s$ is the transmission energy and
$\tilde{\boldsymbol{\Lambda}}_k \in \mathbb{C}^{N_T \times N_R}$ is
a diagonal matrix whose entries are the variances of the estimation
errors.

\subsection{Iterative Detection and Decoding}

\begin{figure}[htb]
\begin{minipage}[h]{0.9\linewidth}
  \centering
  \centerline{\epsfig{figure=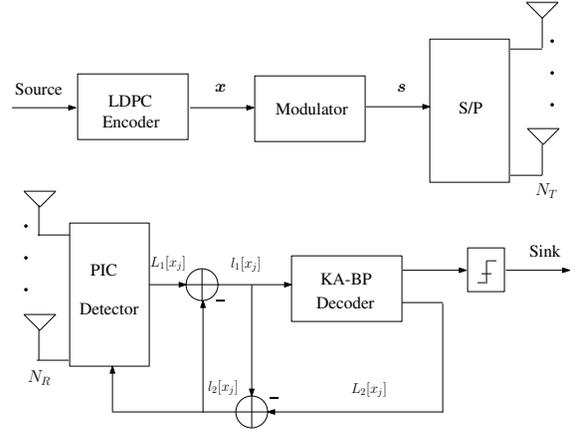,scale=0.375}}
  \vspace{-0em}\caption{Iterative LDPC-coded MIMO spatial multiplexing system with a SISO PIC-MMSE detector
  and the proposed KA-BP decoders.} \label{fig:system}
\end{minipage}
\end{figure}

A block diagram of the IDD system employed in this work is depicted
in Fig. \ref{fig:system}. With the PIC-MMSE processing, we set $y_k
= s_k + {n}_{\scriptsize \mbox{eff}}$ at the output of the detector,
where ${n}_{\scriptsize \mbox{eff}}$ is the effective noise factor
after the MMSE filtering. By assuming that the output of the $k$-th
layer $y_k$ is statistically independent from the other layers
\cite{Wang}, this leads to the approximation of the log-likelihood
ratio (LLR) of bit $x_{k,j}$
\begin{equation}\label{4}
\Lambda_1[x_{k,j}] \approx \log
\frac{P(x_{k,j}=+1|{y_k})}{P(x_{k,j}=-1|{y_k})} =
\lambda_1[x_{k,j}]+\lambda_2^p[x_{k,j}],
\end{equation}
where the last term represents the \textit{a priori} information for
the coded bits $x_{k,j}$, which is obtained by the LDPC decoder. The
first term $\lambda_1$ denotes the \textit{extrinsic} information
which is computed based on $\boldsymbol {r}$ and  the \textit{a
priori} information $\lambda_2^p$. For the detector, by relaxing the
stream index $k$, the coded bit \textit{extrinsic} LLR is obtained
as
\begin{equation}\label{extrif}
\lambda_1[x_{j}]=\log
\frac{\sum_{a_c{\in}\mathcal{A}_{j}^+}P(y|s=a_c)\exp(L_a(a_c))}{\sum_{a_c{\in}\mathcal{A}_{j}^-}P(y|s=a_c)\exp(L_a(a_c))}
\end{equation}
where $\mathcal{A}_{j}^+$ and $\mathcal{A}_{j}^-$ denotes the
subsets of constellation $\mathcal{A}$ where the bit $x_{j}$ takes
the values 1 and 0, respectively. The value $L_a(a_c)$ denotes the
\textit{a priori} symbol probability for symbol $a_c$ and
\begin{equation}
 P(y | s =
a_c) = \frac{1}{\pi\sigma_{\scriptsize
\mbox{eff}}^2}\exp({\frac{-|y-s|^2}{\sigma_{\scriptsize
\mbox{eff}}^2}})
\end{equation}
For an IDD scheme, the computed $\lambda_1$ is fed to the LDPC
decoder as the \textit{a priori} information. The LDPC decoder
calculates the \textit{a posteriori} LLR of each code bit as will be
detailed later.

\section{Knowledge-Aided Decoding Algorithms for IDD Schemes}

The proposed CKAR-BP and EKAR-BP algorithms are designed to improve
the convergence behaviour of the standard BP algorithm by
reweighting part of the hypergraph. These algorithms take the short
cycles into account, such that the decoder can generate more
accurate marginal distributions corresponding to coded data. The
reweighting strategy was first employed in the tree-reweighted BP
(TRW-BP) algorithm reported in \cite{Wainright2}, where the authors
reformulated the BP decoding problem into a tractable convex
optimization problem that iteratively computes beliefs and factor
appearance probabilities (FAPs). Later with the same concept but
additional constraints, the uniformly reweighted BP (URW-BP)
algorithm \cite{Wymeersch2} was introduced for which the FAPs were
constrained to be a constant. A disadvantage of URW-BP is that it
can only be applied to regular LDPC codes. Compared to those two
methods, CKAR-BP and EKAR-BP algorithms optimize the FAPs off-line
by relaxing the constraints from \cite{Wainright2} and
\cite{Wymeersch2}. Additionally, neither of them impose extra
computational complexity to online decoding. Next, we present
general message passing rules for reweighted BP algorithms, then
elaborate both CKAR-BP and EKAR-BP decoders.

\subsection{Message Passing Rules for Knowledge-Aided Decoders}

The message passing rules of reweighted BP algorithms are briefly
reviewed here, the derivation of which can be found in
\cite{Wainright2} with pairwise interactions and in
\cite{Wymeersch2} with higher-order interactions. Given a hypergraph
having $N$ variable nodes and $M$ check nodes and the reweighting
vector $\boldsymbol{\rho}=[\rho_{1},\rho_{2},\ldots,\rho_{M}]$, the
message from the $j$-th variable node $s_j$ to the $i$-th check node
$c_i$ is given by
\begin{equation} \label{mp1}
\centering {\Psi_{ji}=\lambda_{\mathrm{In},j}+\sum_{i'\in \mathcal{N}(j)\backslash
i}\rho_{i'}\Lambda_{i'j}-(1-\rho_i)\Lambda_{ij}},
\end{equation}
where $i'\in \mathcal{N}(j)\backslash i$ is the neighboring set of
check nodes of $s_j$ except $c_i$. Since all messages are
represented in LLRs, $\lambda_{\mathrm{In},j}$ is equal to
$l_1[x_{j}$ in the first decoding iteration. We use $\Lambda_{ij}$
to denote messages sent from $c_i$ to $s_j$ in previous decoding
iterations, then for check nodes $c_i$ $\Lambda_{mn}$ is updated as
\begin{equation} \label{mp2}
\centering {\Lambda_{ij}=2 \mathrm {tanh}^{-1}\big{(}\prod_{j'\in
\mathcal{N}(i)\backslash j}\mathrm{tanh}
\frac{\Psi_{j'i}}{2}}\big{)},
\end{equation}
where `$\rm {tanh}(\cdot)$` denotes the hyperbolic tangent function
as in the standard BP message passing rule to compute an LLR message
from check node $c_i$ to variable node $s_j$. Finally, we have the
belief $b({x_j})$ with respect to $x_j$ given by
\begin{equation} \label{mp3}
\centering
b({x_j})=\lambda_{\mathrm{In},j}+\sum_{i\in \mathcal{N}(j)}\rho_i\Lambda_{ij}.
\end{equation}
The proposed KA-BP decoders iteratively employ
\eqref{mp1}-\eqref{mp3} to update the message regarding each node.
At the end of decoding, $\lambda_{\mathrm{Belief},j}$ serves as the
soft output for deciding the value of $\hat{x}_j$ or for generating
the extrinsic information $l_2[x_j]$ in the next IDD iteration.
Notice that $\rho_i=1, \forall i$ corresponds to the standard BP
algorithm so that no additional complexity is introduced due to the
presence of $\boldsymbol{\rho}$ in real-time decoding.

\subsection {Cycles Knowledge-Aided Reweighted BP (CKAR-BP)}

\begin{table}[!t]
\centering \caption{\label{tab:CKAR-BP}} \vspace{-0.5em} Algorithm
Flow of CKAR-BP Decoder
\begin{small}
\begin{tabular}{ll}
\hline & \tabularnewline \textbf{{Offline Stage 1: counting short
cycles}} & \tabularnewline
 & \tabularnewline
1: Run the algorithm \cite{Halford} to count the number of cycles & \tabularnewline with length-$g$ passing the check node $c_i, \forall i$; & \tabularnewline
 & \tabularnewline
\textbf{{Offline stage 2: determination of $\boldsymbol{\rho}_{i}$ for the hypergraph}} & \tabularnewline
 & \tabularnewline
2: Determine variable FAPs for each check node: & \tabularnewline if
${g}_{C_i}<\mu_g$ $\rho_i=1$, otherwise $\rho_i=\rho_v$ where
$\rho_v=2/\bar{n_D}$;  & \tabularnewline
 & \tabularnewline
\textbf{{Online Stage: real-time decoding }} & \tabularnewline
 & \tabularnewline
3: Update the belief $b(x_j)$ iteratively using reweighted &
\tabularnewline message passing rules \eqref{mp1}--\eqref{mp3} with
optimized & \tabularnewline
$\boldsymbol{\rho}=[\rho_{1},\rho_{2},\ldots,\rho_{M}]$. Decoding
stops if $\boldsymbol{H {\hat x}^T}=\boldsymbol{0}$ or the &
\tabularnewline maximum number of decoding iterations is reached. &
\tabularnewline
& \tabularnewline \hline %& \tabularnewline
\end{tabular}\end{small} \vspace{-1.5em}
\end{table}

Given the knowledge of the distribution of cycles in the graph, the
CKAR-BP algorithm selects the reweighting parameters in order to
mitigate the effect of short cycles, i.e. the statistical dependency
among the incoming messages being exchanged by nodes, leading to a
situation in which the outgoing messages inaccurately have a high
reliability or equivalently a low quality. The algorithm
\cite{Halford}, used for counting short cycles, is a matrix
multiplication technique which can find the girth $g$ implicitly and
calculate the number of cycles with length of $g$, $g+2$ and $g+4$,
explicitly. As shown in Table \ref{tab:CKAR-BP}, after running the
algorithm for counting cycles and calculating $\mu_g$ the average
number of length-$g$ cycles passing a check node, we determine the
reweighting parameters $\rho_i (i=0, 1, \ldots, {M-1})$ under a
simple criterion:
\begin{equation}
\rho_i = \left\{ \begin{array}{ll}
1 &  \textrm{ if ~~ ${g}_{C_i} <  \mu_g$}, \\
\rho_v & {\rm otherwise,}\\
\end{array}\right.
\end{equation}
where ${g}_{C_i}$ is the number of length-g cycles passing a check
node $C_i$, $\rho_v=2/\bar{n_D}$ and $\bar{n_D}$ is the average
connectivity for $N$ variable nodes, which is computed by:
\begin{equation} \label{AC}
\centering {
\bar{n_D}=\frac{1}{\int_{0}^{1}{\upsilon(x)}dx}=\frac{M}{N\int_{0}^{1}{\nu(x)}dx}},
\end{equation}
where $\upsilon(x)$ and $\nu(x)$ are distributions of the variable
nodes and the check nodes, respectively. As an improvement to the
URW-BP algorithm \cite{Wymeersch2}, the proposed CKAR-BP requires
additional complexity due to the cycle counting algorithm
\cite{Halford}. Most importantly, CKAR-BP algorithm can improve the
performance of the BP algorithm when decoding LDPC codes with both
uniform structures (regular codes) and with non-uniform structures
(irregular codes). More details of CKAR-BP and its applications can
be found in \cite{vfap}.

\subsection {Expansion Knowledge-Aided Reweighted BP (EKAR-BP)}
\begin{table}[!t]
\centering \caption{ \vspace{-0.5em} \label{tab:EKAR-BP}}Algorithm
Flow of EKAR-BP Decoder

\begin{small}
\begin{tabular}{ll}
\hline & \tabularnewline

\textbf{{Offline Stage 1: subgraphs formation} } & \tabularnewline
 & \tabularnewline
1: Given a hypergraph $\mathcal{G}$ and $d_{\mathrm{max}}$, apply the modified PEG & \tabularnewline expansion to generate
$T\ge1$ subgraphs; & \tabularnewline
 & \tabularnewline
\textbf{{Offline Stage 2: optimization of $\boldsymbol{\rho}_{t}$ for the
$t$-th subgraph}} & \tabularnewline
 & \tabularnewline
2: Initialize $\boldsymbol{\rho}_{t}^{(0)}$ to a valid value;  & \tabularnewline
 & \tabularnewline
3: For each subgraph, calculate the beliefs $b(\boldsymbol{x}_t)$
and & \tabularnewline the mutual information term
$\boldsymbol{I}_{t}=[I_{t,1},I_{t,2},\ldots,I_{t,L_{t}}]$ &
\tabularnewline by using reweighted message passing rule
\eqref{mp1}--\eqref{mp3}; \\
 & \tabularnewline
4: With $b(\boldsymbol{x}_t)$ and $\boldsymbol{I}_{t}$ obtained from
step 3, update  & \tabularnewline $\boldsymbol{\rho}_{t}^{(r)}$ to
$\boldsymbol{\rho}_{t}^{(r+1)}$ using the conditional gradient
method; & \tabularnewline

 & \tabularnewline
5: Repeat steps 3--4 until $\boldsymbol{\rho}_{t}$ converges for
each subgraph; & \tabularnewline
 & \tabularnewline
\textbf{Offline Stage 3: choice of $\boldsymbol{\rho}=[\rho_{1},\rho_{2},\ldots,\rho_{M}]$ for
decoding}  & \tabularnewline
 & \tabularnewline
6: For all $T$ subgraphs, collect $\boldsymbol{\rho}_{1},\ldots,
\boldsymbol{\rho}_{i}, \ldots, \boldsymbol{\rho}_{T}$.  &
\tabularnewline In case of multiple values $\rho_{i}$ for the same $i$-th check node, & \tabularnewline
choose the one offering
the best performance; & \tabularnewline
 & \tabularnewline
\textbf{{Online Stage: real-time decoding }} & \tabularnewline
 & \tabularnewline
7: Update the belief $b(x_j)$ iteratively using reweighted &
\tabularnewline message passing rules \eqref{mp1}--\eqref{mp3} with
the optimized & \tabularnewline
$\boldsymbol{\rho}=[\rho_{1},\rho_{2},\ldots,\rho_{M}]$. Decoding
stops if $\boldsymbol{H {\hat x}^T}=\boldsymbol{0}$ or the &
\tabularnewline maximum number of decoding iterations is reached. &
\tabularnewline

& \tabularnewline
\hline %& \tabularnewline
\end{tabular}\end{small}\vspace{-1.5em}
\end{table}

The proposed EKAR-BP algorithm transforms the original hypergraph
$\mathcal{G}$ into a set of $T\ge1$ subgraphs and then locally
optimizes the reweighting parameter vector $\boldsymbol{\rho}_t,
t=1,2, \dots, T$ with respect to each subgraph, where the size of
the $t$-th subgraph determines the dimension of
$\boldsymbol{\rho}_t$. It should be noted that $T=1$ corresponds to
the original TRW-BP algorithm \cite{Wainright2} which has a
computational complexity of $\mathcal{O}(M^{2}N)$ and the
convergence of $\boldsymbol{\rho}$ is very slow for large graphs.
Nevertheless, the optimization of $\boldsymbol{\rho}$ could be
significantly less complex when more subgraphs are considered
($\boldsymbol{\rho}$). Thus, there is a need for a flexible method
to decompose the original hypergraph into subgraphs. Inspired by
\cite{Hu}, we apply a modified progressive-edge growth (PEG)
approach to achieve the hypergraph expansion. Generally, the number
of subgraphs $T$ depends on a pre-set maximum expansion level
$d_{\mathrm{max}}$, as a large $d_{\mathrm{max}}$ results in a small
$T$ but a high probability of existence of very short cycles within
subgraphs. Compared to the greedy version of PEG \cite{Hu}, our
modified PEG expansion has two differences: (i) the expansion stops
as soon as every member of the set of nodes $V_t$ has been visited;
(ii) the number of edges incident to the node $s_{j}$ might be less
than its degree since some short cycles are excluded in subgraphs to
guarantee that the local girth of each subgraph $g_t$ is always
larger than the global girth of the original graph $g$.

As shown in Table. \ref{tab:EKAR-BP}, after obtaining $T$ subgraphs,
we introduce $\boldsymbol{L}=[L_{1},L_{2},\ldots,L_{T}]$ in which
$L_{t}$ is the number of check nodes in the $t$-th subgraph. Note
that $\sum_{t}L_{t}>M$ due to duplicated nodes during hypergraph
expansion. With the $t$-th subgraph, we optimize the associated FAPs
$\boldsymbol{\rho}_t=[\rho_{t,1},\rho_{t,2},\ldots,\rho_{t,L_t}]$
using a recursive optimization method, similar to TRW-BP
\cite{Wainright2} but with higher-order interactions and related
message passing rules \eqref{mp1}--\eqref{mp3}. The optimization
problem is solved recursively as follows: 1) for all $T$ subgraphs
in parallel and fixed $\boldsymbol{\rho}_{t}^{(r)}$, use message
passing rules \eqref{mp1}--\eqref{mp3} to calculate the beliefs
$b(\boldsymbol{x}_t)$ as well as the mutual information term
$\boldsymbol{I}_{t}=[I_{t,1},I_{t,2},\ldots,I_{t,L_{t}}]$ provided
with $L_{t}\le M$ check nodes in the $t$-th subgraph; 2) for all $T$
subgraphs in parallel, given $\{\boldsymbol{I}_{t}\}_{t=1}^{T}$, use
the conditional gradient method to update, for all $t$,
$\boldsymbol{\rho}_{t}^{(r)}$ to $\boldsymbol{\rho}_{t}^{(r+1)}$,
then go back to step 1).

The optimization problem is given by
\begin{align*}
\mathrm{minimize} & \,\,\,\,-\boldsymbol {\rho}_t^{\dagger}\boldsymbol{I}_{t}\\
\mathrm{s.t.} & \,\,\,\,\boldsymbol{\rho}_{t} \in \mathbb{T}\big{(}\mathcal{G}_{t}\big{)},
\end{align*}
where $(\cdot)^{\dagger}$ denotes matrix transpose,
$\mathbb{T}\big{(}\mathcal{G}_{t}\big{)}$ is the set of all valid
FAPs over the subgraph $\mathcal{G}_{t}$ and $I_{t,l}$ is a mutual
information term depending on $\boldsymbol {\rho}^{(r)}_t$, the
previous value of $\boldsymbol {\rho}_t$. By denoting the objective
function as $f(\boldsymbol {\rho}_t)=-\boldsymbol
{\rho}_t^{\dagger}\boldsymbol{I}_{t}$, we first linearize the
objective around the current value $\boldsymbol {\rho}^{(r)}_t$:
\begin{equation}\label{flin}
f_{\mathrm{lin}}(\boldsymbol {\rho}_t)=f(\boldsymbol {\rho}^{(r)}_t) + \nabla_{\boldsymbol {\rho}_t}^{\dagger}f(\boldsymbol {\rho}^{(r)}_t) (\boldsymbol {\rho}_t-\boldsymbol {\rho}^{(r)}_t),
\end{equation}
where$\nabla_{\boldsymbol {\rho}_t}f(\boldsymbol
{\rho}^{(r)}_t)=-\boldsymbol{I}_{t}$. Secondly, we minimize
$f_{\mathrm{lin}}(\boldsymbol {\rho}_t)$ with respect to
$\boldsymbol {\rho}_t$, denoting the minimizer by
$\boldsymbol{\rho}_t^{\ast}$ and
$z^{(r+1)}_t=\max(f_{\mathrm{lin}}(\boldsymbol{\rho}_t^{\ast}),z^{(r)}_t)$,
where $z^{0}_t=-\infty$.  Finally, $\boldsymbol {\rho}^{(r)}_t$ is
updated to $\boldsymbol {\rho}^{(r+1)}_t$ as
\begin{equation}\label{11}
\centering{\boldsymbol{\rho}_t^{(r+1)}=\boldsymbol{\rho}_t^{(r)}+\alpha(\boldsymbol{\rho}_t^{\ast}-\boldsymbol{\rho}_t^{(r)})},
\end{equation} in which $\alpha$ is chosen as
\begin{equation} \label{10}
\arg \min_{\alpha \in [0,1]}f(\boldsymbol{\rho}_t^{(r)}+\alpha(\boldsymbol{\rho}_t^{\ast}-\boldsymbol{\rho}_t^{(r)})).
\end{equation}
At every recursion, $f(\boldsymbol{\rho}_t^{(r)})$ is an upper bound
on the optimized objective, while $z_t^{(r+1)}$ is a lower bound.
Note that the proposed EKAR-BP algorithm is straightforward to use
if the LDPC code was designed by PEG, or its variations
\cite{uchoa,con}, but is not limited to such designs.

\section{Simulation Results}

In this section, we present the simulation results of the proposed
IDD scheme with the CKAR-BP and EKAR-BP algorithms for a $4 \times
4$ LDPC-coded MIMO system with PIC-MMSE detection.  The LDPC code is
a regular code designed by the PEG algorithm \cite{Hu} whose block
length $N$ is $1000$, the rate $R$ is $0.5$, the girth $(g)$ is $6$,
and the degree distributions are $3 (\upsilon(x)=x^4)$ and $5
(\nu(x)=x^6)$ respectively. We consider uncorrelated Rayleigh flat
fading channels and used $30$ inner decoding iterations in this
experiment. For the EKAR-BP decoder, $T=20$ subgraphs have been
generated, where check nodes are allowed to be re-visited, and $600$
recursions were employed to obtain ${\boldsymbol \rho}$.

\begin{figure}[!htb]
\begin{center}
\def\epsfsize#1#2{0.9\columnwidth}
\epsfbox{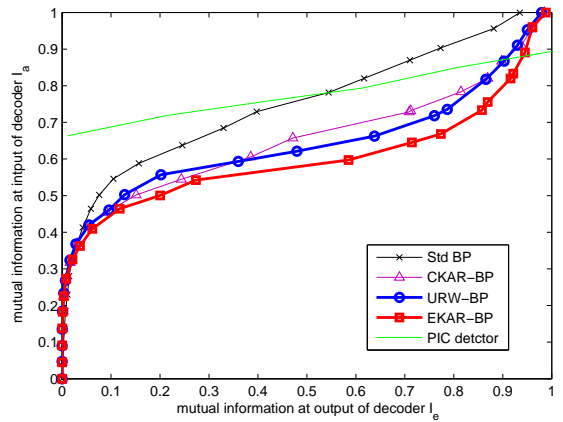} \vspace{-1.25em}  \caption{EXIT charts of
different decoders with a PIC detector. The proposed EKAR-BP decoder
matches better with the PIC detector than other decoders. The EXIT
chart of the PIC detector is obtained at
$E_b/N_0=4$dB.}\label{fig:exit}
\end{center}
\end{figure}

\begin{figure}[!htb]
\begin{center}
\def\epsfsize#1#2{0.9\columnwidth}
\epsfbox{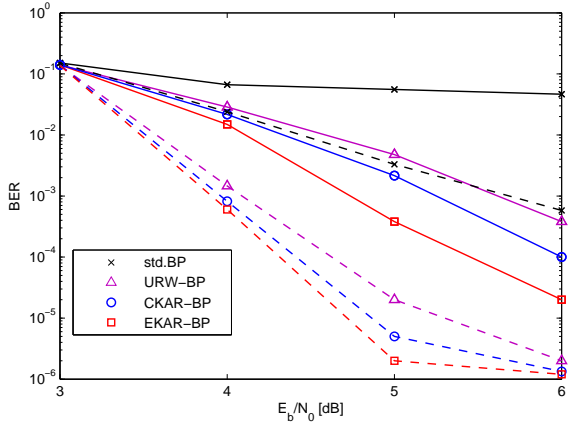} \vspace{-1.25em}\caption{Comparison of the
standard BP, URW-BP, CKAR-BP, and EKAR-BP in terms of BER
performances for a $4 \times 4$ system. }\label{fig:ber}
\end{center}
\end{figure}

In comparison with the standard BP and URW-BP algorithms, we first
draw an \textit{extrinsic} information transfer (EXIT) charts of
different decoders with the SISO PIC detector in Fig.
\ref{fig:exit}. Although the curve of the PIC-MMSE detector does not
reach the top-right $(1,1)$ point at the given SNR, it is obvious
that the combination of PIC-MMSE detector and the proposed EKAR-BP
decoder creates the widest detection and decoding tunnel.
Additionally, only the tunnel between the PIC-MMSE detector and the
standard BP decoder is closed at an early stage, which indicates
that performance gain from the IDD process could be significantly
diminished in this case. To verify the result of the EXIT chart,
Fig. \ref{fig:ber} depicts the performance in bit-error ratio (BER)
of the MIMO system. We have used $30$ inner decoding iterations and
up to $3$ outer detection and decoding iterations. The performance
curves after $2$ outer iterations are denoted by solid lines while
the curves after $3$ outer iterations are denoted by dashed lines.
From Fig. \ref{fig:ber}, both CKAR-BP and EKAR-BP decoders
outperform the standard BP and URW-BP decoder in the first detection
and decoding iteration. In the third outer iteration, the proposed
decoders are still able to generate relatively good performance when
considering the low SNR range and the block length of code. Notice
that there is an error floor effect at the BER of $10^6$, which can
be mitigated by using decision feedback techniques,
\cite{delamare_spadf}, \cite{Peng} and \cite{Peng2}. As mentioned in
Section \uppercase\expandafter{\romannumeral 3}, the key feature of
the proposed KA-BP decoders lies in that no additional complexity is
imposed in real-time decoding since the optimization of
$\boldsymbol{\rho}$ is carried out offline. Moreover, by increasing
the number of subgraphs $T$ the EKAR-BP can accelerate the
optimization process such that it can be employed for time-varying
channels.

\section{conclusion}
We have proposed an IDD scheme for MIMO systems with a conventional
PIC-MMSE detector and two novel KA-BP decoders, which implement the
reweighting strategy for decoding finite-length regular or irregular
LDPC codes. The proposed CKAR-BP and EKAR-BP algorithms have
different computational costs in the optimization phase, but neither
of which requires extra complexity for online decoding. Furthermore,
the EKAR-BP algorithm provides a trade-off between the number of
expanded subgraphs and the convergence speed of the reweighting
parameters. Numerical results show that the proposed IDD system is
able to offer good performance while using a reduced number of inner
and outer iterations.

\end{document}